\documentclass{article}

\usepackage{microtype}
\usepackage{graphicx}
\usepackage{subfigure}
\usepackage{booktabs} 

\usepackage{hyperref}

\usepackage[accepted]{icml2024}

\usepackage{amsmath}
\usepackage{amssymb}
\usepackage{mathtools}
\usepackage{amsthm}

\usepackage[capitalize,noabbrev]{cleveref}

\theoremstyle{plain}

\theoremstyle{definition}

\theoremstyle{remark}

\usepackage[textsize=tiny]{todonotes}

\icmltitlerunning{Transformer-Based Astronomical Time Series Model with Uncertainty Estimation for Detecting Misclassified Instances}

\begin{document}

\twocolumn[
\icmltitle{ Transformer-Based Astronomical Time Series Model with Uncertainty Estimation for Detecting Misclassified Instances}



\icmlsetsymbol{equal}{*}

\begin{icmlauthorlist}
\icmlauthor{Martina Cádiz-Leyton}{yyy}
\icmlauthor{Guillermo Cabrera-Vives}{yyy,comp,tt,dos}
\icmlauthor{Pavlos Protopapas}{sch}
\icmlauthor{Daniel Moreno-Cartagena}{yyy,comp}
\icmlauthor{Cristobal Donoso-Oliva}{comp,tt}


\end{icmlauthorlist}

\icmlaffiliation{yyy}{ Department of Computer Science, Universidad de Concepción, Chile}
\icmlaffiliation{comp}{Center for Data and Artificial Intelligence, Universidad de Concepción, Edmundo Larenas 310, Concepción, Chile}
\icmlaffiliation{sch}{John A. Paulson School of Engineering and Applied Science, Harvard University, Cambridge, MA, 02138}
\icmlaffiliation{dos}{Millennium Institute of Astrophysics (MAS), Nuncio Monseñor Sotero Sanz 100, Of. 104, Providencia, Santiago, Chile}
\icmlaffiliation{tt}{Millennium Nucleus on Young Exoplanets and their Moons (YEMS), Chile}

\icmlcorrespondingauthor{Martina Cádiz-Leyton}{mcadiz2018@inf.udec.cl}
\icmlcorrespondingauthor{Guillermo Cabrera-Vives}{guillecabrera@inf.udec.cl}

\icmlkeywords{Machine Learning, ICML}

\vskip 0.3in
]



\printAffiliationsAndNotice{} 

\begin{abstract}

In this work, we present a framework for estimating and evaluating uncertainty in deep-attention-based classifiers for light curves for variable stars. We implemented three techniques, Deep Ensembles (DEs), Monte Carlo Dropout (MCD) and Hierarchical Stochastic Attention (HSA) and evaluated models trained on three astronomical surveys. Our results demonstrate that MCD and HSA offers a competitive and computationally less expensive alternative to DE, allowing the training of transformers with the ability to estimate uncertainties for large-scale light curve datasets. We conclude that the quality of the uncertainty estimation is evaluated using the ROC AUC metric.
\end{abstract}

\section{Introduction}
\label{sec:Introduction}

The identification and classification of variable stars are critical for advancing our understanding of the cosmos. For example, Cepheids and RR Lyrae stars are key astronomical distance indicators, aiding in measuring the scale of the universe \cite{feast2014cepheid, ngeow2015application}. The emergence of wide-field surveys, such as the Large Synoptic Survey Telescope (LSST; \citealt{ivezic2019lsst}), which will generate about 20 TB of data nightly, provides unprecedented opportunities for the analysis of photometric observations. Such comprehensive data, with their corresponding time-stamps, are instrumental in detecting new classes of variable stars and uncovering previously unknown astronomical phenomena \cite{bassi2021classification}.
    
Astronomical time series classification, essential for variable star analysis, faces challenges due heteroskedasticity, sparsity, and observational gaps \cite{mahabal2017deep}. Astronomy has shifted from traditional feature-based analysis to sophisticated data-driven models through deep learning \cite{smith2023astronomia}, transitioning to architectures from multilayer perceptrons \cite{karpenka2013simple} to Recurrent Neural Networks (RNNs) and Convolutional Neural Networks (CNNs), showcasing significant advancements in the field \cite{cabrerareyes, mahabal2017deep, protopapas2017recurrent, naul2018recurrent, carrasco2019deep, becker2020scalable, donoso2021effect}. 
    
Building upon this evolution, self-attention-based models \cite{vaswani2017attention} like ASTROMER \cite{donoso2022astromer}-a transformer-based embedding model for learning single-band astronomical time series representations-have set new benchmarks in the field \cite{moreno2023positional, pan2023astroconformer, leung2023towards, tanoglidis2023transformers}. These models have broadened applications from denoising time series \cite{morvan2022don} to specialized multi-band Supernovae classification \cite{pimentel2022deep}. 
    
As the complexity of models increases, ensuring the trustworthiness of classification outcomes become a critical concern. In the context of deep neural networks, uncertainty estimation plays a crucial role, enhancing prediction confidence and is particularly useful in misclassification detection, where uncertain predictions can flag classification errors \cite{gawlikowski2023survey}. Traditional methods, such as Bayesian Neural Networks (BNNs; \citealt{blundell2015weight}), which approximate a posterior distribution over model parameters, have proven effective in time-domain astronomy \cite{moller2020supernnova}. However, they are inherently computationally demanding.

In this work, we present a framework that combines deep attention-based classifiers with uncertainty estimation techniques for misclassification detection. Our model utilizes the ASTROMER transformer-based embedding, supported by the Chilean-led ALeRCE astronomical alert broker \cite{forster2021automatic}. We incorporate several methods of uncertainty estimation, including Deep Ensembles (DEs; \citealt{lakshminarayanan2017simple, ganaie2022ensemble}), Monte Carlo Dropout (MCD; \citealt{gal2016dropout}), and Hierarchical Stochastic Attention (HSA; \citealt{pei2022transformer}), each applied to the ASTROMER framework. Although DEs require significant computational resources, they establish a robust baseline for capturing uncertainty and facilitate comparisons with other techniques. Our empirical evaluations indicate that MCD matches performance of DEs, while HSA demonstrates superior predictive performance. 
      
\section{Methodology}
\label{sec:methods}

\subsection{Transformer-based Model Architecture}
\label{subsec:model_arch}
Our work utilizes ASTROMER \cite{donoso2022astromer}, a transformer architecture inspired by BERT \cite{devlin2019google}. Unlike BERT, which is designed for Natural Language Processing (NLP) tasks, ASTROMER is specifically adapted for analyzing astronomical time series. The model includes several key components: positional encoding (PE), two self-attention blocks, and a long-short-term memory network (LSTMs; \citealt{hochreiter1997long}) classifier.

The model processes single-band time series with $L$ observations, each defined by a vector of magnitudes $x \in \mathbb{R}^L$ and corresponding timestamps $t \in \mathbb{R}^L$. The PE component projects each timestamp into an embedding space $X \in \mathbb{R}^{L \times d_x}$, with $d_x = 256$, which combines with magnitude data through a feedforward network to produce a $200 \times 256$ matrix input for the self-attention mechanism.

The self-attention block transforms $X$ into query \(Q\), key \(K\), and value \(V\) matrices using trainable weights \(W_Q, W_K, W_V\), and computes the attention output as:
\begin{equation}
Q = W_Q X; \quad K = W_K X; \quad V = W_V X;
\end{equation}
\begin{equation}
A = \text{softmax}\left(\frac{QK^T}{\sqrt{d_k}}\right); \quad H = AV.
\label{eq:attention}
\end{equation}

Here, \( A = [a_1,.., a_h]\)  is the attention distribution containing normalized weights, and \( H = [h_1,.., h_h]\) is the attention output. This setup includes four attention heads, each with 64 neurons, and integrates dropout layers with a rate of 0.1 to prevent overfitting. The output from this transformer feeds into a two-layer LSTM network, which culminates in a softmax layer for classification.

We initialized our classifiers with pre-trained weights from 1,529,386 R-band time-magnitude light curves of the MACHO survey \cite{alcock2000macho}, applying uncertainty quantification techniques to enhance reliability in variable star classification.

\subsection{Deep Ensembles}
\label{subsec:deep_ens}

DEs offer a practical non-Bayesian alternative to BNNs for modeling predictive uncertainty without the computational complexity associated with BNNs. DEs utilize multiple, independently trained neural networks to approximate the predictive distribution for a new data point $x^*$ as:

\begin{equation}
    y^* \approx \frac{1}{T} \sum_{t=1}^T f_{\theta_t}(x^*), \label{eq:MC}
\end{equation}
where $\{\theta_t\}_{t=1}^T$ are the parameters of $T$ decorrelated neural networks. This approach mitigates the intractable integral over the predictive distribution:

\begin{equation}
    p(y^*|x^*, \mathcal{D}) = \int p(y^*|x^*, \theta)p(\theta|\mathcal{D})d\theta. \label{eq:predictive_distribution}
\end{equation}

To calculate statistics over our uncertainty estimates, we trained ten ensembles, each consisting of ten models, for each survey, using different random seeds. This approach enables robust uncertainty estimation by averaging outputs from various models tested on the same dataset.

\subsection{Monte Carlo Dropout}
Dropout was originally designed to prevent overfitting \cite{srivastava2014dropout}. MCD, introduced by \cite{gal2016dropout}, extends this concept by applying dropout during inference, thus approximating the predictive distribution as described in Equation~\ref{eq:predictive_distribution}. This adaptation enables MCD to simulate an ensemble of diverse models. Each stochastic pass deactivates a random subset of neurons, thereby enhancing the robustness and reliability of predictions by capturing a broader spectrum of network states.

The expected value of the output \( y^* \) for a given input \( x^* \) is computed by averaging the outputs from the model \( f_{\theta_t}(x^*) \) across all dropout samples, as shown in Equation~\ref{eq:MC}. In our adaptation of the ASTROMER model, we have implemented the “MCD All” methodology as proposed by Shelmanov et al. \cite{shelmanov2021certain}, which applies Monte Carlo Dropout uniformly across all dropout layers of the attention blocks to better capture model uncertainty.

\subsection{Hierarchical Stochastic Attention}
\label{subsec:hsa}

HSA \cite{pei2022transformer} technique enhances the self-attention layers by incorporating stochasticity through the Gumbel-softmax distribution and learnable centroids within the attention block. The procedure allows for differentiable sampling of attention weights, maintaining the capacity for gradient-based optimization while introducing variability. 

It modifies the standard multi-head attention by using learnable centroids \(C \in \mathbb{R}^{d_k \times c}\), each corresponding to a key vector, and employing the Gumbel-softmax distribution:
\begin{equation}
    \mathcal{G}(\alpha, \tau) = \text{Softmax}\left((\log(\alpha) + s)\tau^{-1}\right),
\end{equation}
where \( \alpha \) denotes class probabilities, \( \tau \) is the temperature, and \( s \) are i.i.d samples from Gumbel(0, 1) \cite{gumbel1954statistical}.
 
For a single \(i\)-th head, let \(q_i \in Q\), \(k_i \in K\), \(v_i \in V\), HSA core operation is defined as:
\begin{align}
\hat{a}_c &\sim \mathcal{G}\left(\frac{k_i^T C}{\tau_1}\right), \\
\tilde{k}_i &= \hat{a}_c^TC, \\
\hat{a}_v &\sim \mathcal{G}\left(\frac{q_i \tilde{k}_i^T}{\tau_2}\right), \\
h_i &= \hat{a}_v v_i.
\end{align}

where $\hat{a}_c$ and $\hat{a}_v$ are the attention distributions for centroids and values respectively. These equations detail how the attention weights are recalibrated using a stochastic process that combines centroid influences with query inputs. This method is integrated into the ASTROMER attention mechanism and captures uncertainties through multiple inference iterations, similar to the MCD method.

\subsection{Uncertainty Estimates} 
\label{subsec:uncertainty_estimates}
When a model is trained using only a maximum likelihood approach, the softmax activation function is prone to generate overconfident predictions \citep{guo2017calibration}. To explore and quantify the uncertainty inherent in our model's predictions, we employ uncertainty estimates (UEs). These estimates are not designed to assess the quality of the uncertainty quantification but to measure the extent and variability of uncertainty itself. 

For MCD and HSA, we calculate UEs by performing $T$ forward pass inference runs, where $T$ is a hyperparameter that also represents the number of models in the DEs case. We compute the following UEs: 

\begin{itemize}
    \item Sampled Maximum Probability (\(\mathrm{SMP}\)):
        \begin{equation}
             1 - \max_{c \in C} \overline{p}(y = c | x),
        \end{equation}
        where $\overline{p}(y = c | x) = \frac{1}{T} \sum_{t = 1}^{T} p_{t}(y = c | x)$ is the average probability of class \(c\) over \(T\) inference runs.

    \item Probability Variance (\(\mathrm{PV}\); \citealt{gal2017deep}):
        \begin{equation}
            \frac{1}{C} \sum_{c = 1}^{C} \left( \frac{1}{T} \sum_{t = 1}^{T} \left( p_{t}(y = c | x) - \overline{p}(y = c | x) \right)^2 \right),
        \end{equation}
        reflecting the variance across all classes \(C\).

    \item Bayesian Active Learning by Disagreement (\(\mathrm{BALD}\); \citealt{houlsby2011bayesian}):
        \begin{equation}
            -\sum_{c = 1}^{C} \overline{p^{c}} \log(\overline{p^{c}}) + \frac{1}{T} \sum_{t = 1}^{T} \sum_{c = 1}^{C} p_{t}^{c} \log(p_{t}^{c}),
        \end{equation}
        where \(\overline{p^{c}}\) and \(p_{t}^{c}\) are the mean and individual class probabilities, respectively. This entropy-based metric measures total uncertainty.
\end{itemize}

\subsection{Misclassification instances}
All our models were trained as multiclass classifiers. However, to evaluate the UE scores, we focused on detecting misclassifications. Therefore, during the testing stage, we transformed the task from a multiclass to a binary problem. This approach is similar to the work done by Shelmanov et al. (\citeyear{shelmanov2021certain}) and Vazhentsev et al. (\citeyear{vazhentsev2022uncertainty}) for NLP tasks. New instances $ \tilde e_i$ are computed as:
    \begin{equation}
     {\tilde{e}_i} =
    \begin{cases}
        1, &  y_i \neq  \hat{y}_i, \\
        0, &  y_i =  \hat{y}_i, 
    \end{cases}
    \end{equation}
where $y_i$ is the true label, and $\hat{y}_i$ is the original predicted label. The new instances $\tilde{e}_i$ indicate whether the model made a mistake in predicting the label of the variable source. 
We calculated the ROC AUC scores using the new instances $\tilde {e}_i$ and their corresponding UEs scores. 
After computing the samples, we used the Wilcoxon–Mann–Whitney (WMW) test to assess the significance of our results.

\section{Experimental Setup}
\label{subsec:datadesc}
In this study, we tested our models on three labeled surveys of variable stars: the Optical Gravitational Lensing Experiment (OGLE-III; \citealt{udalski2004optical}), the Asteroid Terrestrial-impact Last Alert System (ATLAS; \citealt{heinze2018first}), and MACHO dataset. The classification scheme follows the methodology established by Becker et al. (\citeyear{becker_2020}), using data observed through different spectral filters.

To simulate limited data scenarios, we used 500 training samples and 100 test samples per class, setting aside 30\% of the training set for validation. This setup ensured that each model was evaluated under consistent conditions, avoiding the mixing of test sets from different catalogs.

We employed the Adam optimizer \cite{kingma2014adam} with a learning rate of $10^{-3}$, a batch size of 512, and Early Stopping based on validation loss after 20 epochs. These hyperparameters were applied across all experiments, facilitating robust significance testing based on ten predictions per model approach.

\section{Results \& Discussion} 
\label{sec:results}

\begin{table}[t]
\caption{Summary of multiclass inference scores (\%) on the test sets for various datasets. Each model is evaluated with respect to the macro F1 and accuracy score.}
\vskip -0.3in
\begin{center}
\begin{sc}
\small
\begin{tabular}{lllllll}
\toprule
Method & Score & MACHO & ATLAS & OGLE-III \\ 
\midrule
{Baseline} & F1  & 68.6{$\pm$}1.7 & 77.8{$\pm$}2.6 & 67.3{$\pm$}2.9 \\
                          & Acc  & 69.5{$\pm$}1.6 & 77.7{$\pm$}2.5 & 68.5{$\pm$}2.9 \\
\midrule
{MCD} & F1 & 68.0{$\pm$}0.6 & 78.8{$\pm$}0.7 & 66.8{$\pm$}1.6 \\
                    All     & Acc  & 69.3{$\pm$}0.5 & 78.8{$\pm$}0.7 & 68.4{$\pm$}1.4 \\
\midrule
{HSA} & F1 & 74.6{$\pm$}1.8 & 82.1{$\pm$}2.2 & 75.7{$\pm$}1.9 \\
                          & Acc  & 75.3{$\pm$}1.8 & 82.2{$\pm$}2.1 & 76.7{$\pm$}1.7 \\
\bottomrule
\end{tabular}
\end{sc}
\end{center}
\label{tab:updated_results}
\vskip -0.2in

\end{table}

Our evaluation across MACHO, ATLAS, and OGLE-III test sets highlights significant predictive performance variations, with detailed metrics presented in Table \ref{tab:updated_results} focusing on F1 scores and accuracy. The baseline model establishes a consistent benchmark, achieving a notable F1 score and accuracy of around 77.8\% on the ATLAS dataset. The MCD All method exhibits a similar performance to the baseline, with a tendency towards a slight degradation, which is evident in the OGLE-III dataset, where it scores approximately 61.0/63.3 in F1 and accuracy respectively. In contrast, HSA demonstrates a robust performance with an improved F1 score and accuracy, reaching the highest score of 82.1\% on the ATLAS test set. The results indicates that all methods hold the potential for enhanced predictive reliability and consistency across diverse astronomical datasets.

We compare the performance of the methods in terms of the predictive uncertainty for the misclassification detection task on the test set as shown in Table \ref{tab:tabla2_updated}. The evaluation metric used is the absolute ROC AUC score for the DE baseline, while the values for the other techniques represent the difference in performance compared to the corresponding UEs of the baseline.

\begin{table}[t]
\caption{UEs performance in the misclassification task. Mean ROC AUC \(\pm\) standard deviation (\%) scores for each method across the test sets. Statistically significant improvements (p-values \(\leq\) 0.05) compared to the corresponding UE of the baseline are highlighted. Results were assessed using the Wilcoxon-Mann-Whitney (WMW) test to determine statistical significance.}
\label{tab:tabla2_updated}
\vskip 0.1in
\begin{center}
\begin{small}
\begin{sc}
\begin{tabular}{lcccr}
\toprule
\multicolumn{2}{c}{Model} & \multicolumn{3}{c}{Dataset} \\\midrule
Method        & UEs  & MACHO  & ATLAS   & OGLE-III    \\ \midrule
{Baseline}         
& SMP  & 75.6{$\pm$}1.6&85.9{$\pm$}2.1&82.2{$\pm$}1.1\\
& PV   & 71.4{$\pm$}2.3&82.0{$\pm$}2.2&73.4{$\pm$}1.7\\
& BALD & 70.0{$\pm$}2.6&80.8{$\pm$}2.4&74.1{$\pm$}1.8\\ \midrule                
{\begin{tabular}[c]{@{}l@{}}MCD\end{tabular}}  
& SMP  & 0.3{$\pm$}2.0&0.2{$\pm$}1.5&0.0{$\pm$}1.0  \\ 
All& PV   &\textbf{1.4{$\pm$}2.4}&\textbf{1.5{$\pm$}2.3}&\textbf{4.3{$\pm$}1.6}\\ 
& BALD &\textbf{1.8{$\pm$}2.4}&\textbf{1.8{$\pm$}2.6}&\textbf{4.2{$\pm$}1.6}\\  \midrule 

{\begin{tabular}[c]{@{}l@{}}HSA\end{tabular}} 
& SMP  &-1.5{$\pm$}1.8&-1.2{$\pm$}2.1&-0.1{$\pm$}1.0 \\ 
& PV   &0.5{$\pm$}2.4&\textbf{2.6{$\pm$}2.2}&\textbf{5.9{$\pm$}1.5}\\
& BALD &-0.3{$\pm$}3.1&\textbf{2.3{$\pm$}2.5}&\textbf{2.9{$\pm$}1.3} \\  
\bottomrule
\end{tabular}
\end{sc}
\end{small}
\end{center}
\vskip -0.2in
\end{table}

\begin{figure}[ht]
\vskip 0.2in
\begin{center}
\centerline{\includegraphics[width=200pt]{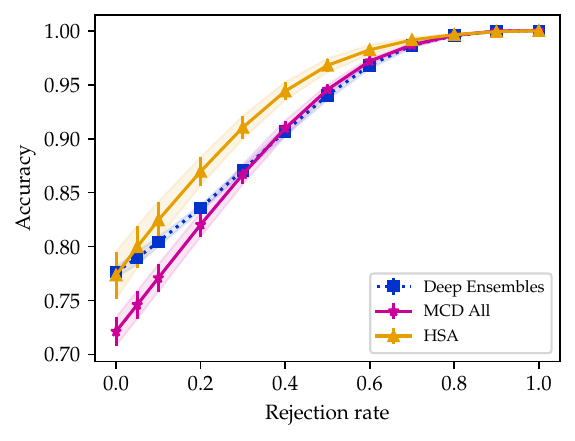}}
\caption{Median accuracy values of the rejection curves for the baseline and the proposed approaches (HSA and MCD All) evaluated on OGLE-III dataset. Each curve is based on the BALD score and binary classification, and the shadings correspond to standard deviation of multiple experiments conducted.}
\label{ogle_plot}
\end{center}
\vskip -0.5in
\end{figure}

The DEs are a strong baseline achieving a mean ROC AUC around 70\% for all UEs and datasets, indicating its ability in identifying potential errors directly from its probabilistic outputs. For PV and BALD UEs, most of the methods outperform the baseline, with the largest increase being a 5.9\% improvement in ROC AUC.  Although we do not observe a statistically significant improvement for SMP, the results do not vary significantly from the baseline. Therefore, the proposed models remain competitive, trailing the baseline only slightly. The following results showcase their adeptness in capturing predictive uncertainty, as seen by their enhanced and close-to-baseline performances in the misclassification task.

To present a practical application of our misclassification framework, we employed the accuracy-rejection plot, as shown in Figure \ref{ogle_plot}. This plot, generated using the BALD score, simulates a scenario that mirrors hybrid machine-human behavior, wherein the machine abstains from classifying the most uncertain samples. As a guideline, to achieve a desired accuracy threshold of 90\%, using the HSA approach, a rejection rate of approximately $\sim 0.35$ is recommended, which implies that the oracle should label about $\sim 0.35$ of the samples. The performance curves of the HSA approach suggest it as a viable alternative to DEs, while MCD maintains similar performance to the baseline accuracy after a rejection rate of $\sim 0.2$.

\section{Conclusions}
\label{sec:conclusions}

We integrated uncertainty estimation techniques DEs, MCD, and HSA into a pre-trained astronomical time series transformer, testing them on MACHO, ATLAS, and OGLE datasets. Results show that these inference-only methods enhance performance efficiently in variable star classification. The adaptability of these techniques to existing architectures demonstrates significant potential for advancing astronomical data analysis.

\section{Acknowledgments}

The authors acknowledge support from the National Agency for Research and Development (ANID) grants: FONDECYT regular 1231877 (MCL, GCV, DMC); Millennium Science Initiative Program – NCN2021\_080 (GCV, CDO) and ICN12 009 (GCV).

\nocite{langley00}

\bibliography{example_paper}
\bibliographystyle{icml2024}

\newpage

\end{document}